\documentclass[times, twoside]{StyleBioRxiv}

\usepackage{xcolor}
\usepackage[nice]{nicefrac}

\renewcommand{\vec}[1]{\mathbf{#1}}
\newcommand{\tensor}[1]{\mathbf{#1}}
\newcommand{\integ}[3]{\int\limits_{#1}^{#2}\!\!\mathrm{d}{#3}\;}
\newcommand{\summe}[3]{\sum\limits_{#1 = #2 }^{#3}\;}

\newcommand{\Pftnull}{\hat{P}_{0,k}}
\newcommand{\rhohom}{\bar{\rho}}
\newcommand{\rhonempol}{\rho^c_\text{nem-pol}}
\newcommand{\rhonempolhydro}{\rho^{(c,h)}_\text{nem-pol}}

\newcommand{\e}{\mathrm{e}}
\newcommand{\imag}{\mathrm{i}}
\newcommand{\Real}{\mathrm{Re}}
\newcommand{\Imag}{\mathrm{Im}}

\newcommand{\polarbias}{\psi}

\leadauthor{Denk} 

\begin{document}

\title{Pattern-induced local symmetry breaking in active matter systems}

\shorttitle{\today}

\author[a,b,c]{Jonas Denk}
\author[a,*]{Erwin Frey}

\affil[a]{Arnold Sommerfeld Center for Theoretical Physics (ASC) and Center for NanoScience (CeNS), Department of Physics, Ludwig-Maximilians-Universit\"at M\"unchen, Theresienstrasse 37, D-80333 M\"unchen, Germany}
\affil[b]{Department of Physics, University of California, Berkeley, CA 94720, USA}
\affil[c]{Department of Integrative Biology, University of California, Berkeley, CA 94720, USA}

\maketitle

\begin{abstract}

The emergence of macroscopic order and patterns is a central paradigm in systems of (self-)propelled agents, and a key component in the structuring of many biological systems.
The relationships between the ordering process and the underlying microscopic interactions have been extensively explored both experimentally and theoretically. 
While emerging patterns often show one specific symmetry (e.g. nematic lane patterns or polarized traveling flocks), depending on the symmetry of the alignment interactions patterns with different symmetries can apparently coexist. 
Indeed, recent experiments with an actomysin motility assay suggest that polar and nematic patterns of actin filaments can interact and dynamically transform into each other. 
However, theoretical understanding of the mechanism responsible remains elusive. 
Here, we present a kinetic approach complemented by a hydrodynamic theory for agents with mixed alignment symmetries, which captures the experimentally observed phenomenology and provides a theoretical explanation for the coexistence and interaction of patterns with different symmetries.
We show that local, pattern-induced symmetry breaking can account for dynamically coexisting patterns with different symmetries.
Specifically, in a regime with moderate densities and a weak polar bias in the alignment interaction, nematic bands show a local symmetry-breaking instability within their high-density core region, which induces the formation of polar waves along the bands. These instabilities eventually result in a self-organized system of nematic bands and polar waves that dynamically transform into each other.   
Our study reveals a mutual feedback mechanism between pattern formation and local symmetry breaking in active matter that has interesting consequences for structure formation in biological systems.
\end {abstract}


\begin{corrauthor}
frey@lmu.de
\end{corrauthor}

Any theory for systems of (self-)propelled agents must be based on assumptions regarding the agents’ propulsion mechanism as well as their interactions.
One of the central insights in active matter theories is that interactions that align the agents' orientations---even if they are short-ranged---can lead to the formation of macroscopic order already in dilute systems in two dimensions~\cite{Ramaswamy2010, Marchetti2013, Vicsek2012, Chate2020}.
Close to the onset of macroscopic order, both experiments with (self-)propelled agents and theoretical studies quite generally observe phase separation into high-density ordered clusters and a low-density disordered background, rather than spatially uniform long-range order~\cite{Vicsek2012,Chate2020,Baer2020}. 
Hence, symmetry breaking in active matter systems seems to be inextricably linked to formation of patterns.

\begin{figure}[t]
\centering
\includegraphics[width=1.\linewidth]{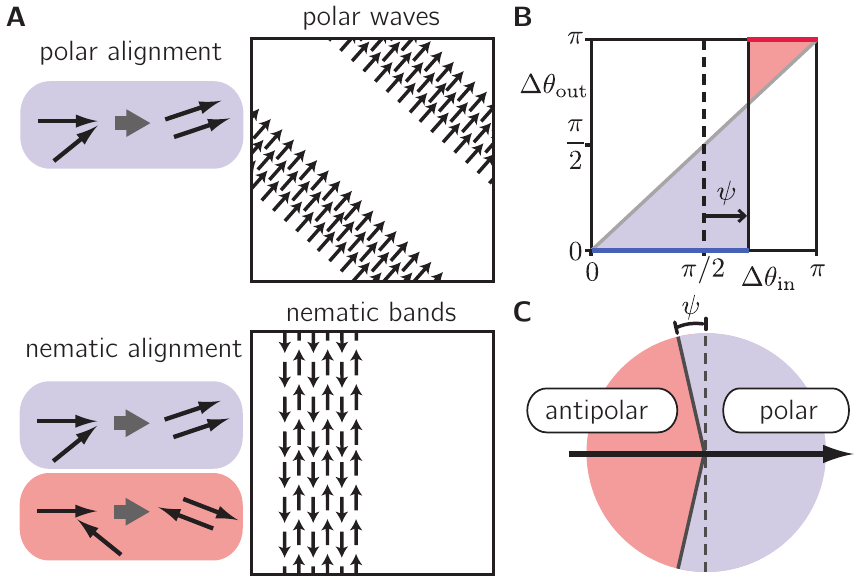}
\caption{\textbf{Symmetries in active matter.} \textbf{(A)} In models with fully polar alignment, polar agents assume the same propulsion direction upon alignment. For fully nematic alignment, particles assume the same or opposing propulsion directions, depending on whether they collide at an acute or obtuse angle, respectively. Polar (nematic) alignment interactions enable macroscopic polar (nematic) order. Beyond the onset of order, the system evolves into wave patterns or nematic bands depending on the symmetry of the alignment interaction.  \textbf{(B)} ‘Collision statistics’ for the binary collision rule with polar bias $\polarbias$. $\Delta\theta_\text{out}$ denotes the angle differences between pairwise particle velocities after the collision and is either $0$ or $\pi$ depending on whether the agents' angle difference before the collision, $\Delta\theta_\text{in}$, is smaller or larger than $\pi/2+\polarbias$, respectively. \textbf{(C)} Illustration of our generalized collision assumption. The black arrow indicates the pre-collision orientation of a reference polar agent. Alignment with a second agent is polar if the propulsion of the second particle lies in the blue shaded angular range and antipolar in the red shaded angular range.}
\label{fig:Symmetries_Model}
\end{figure}

In theoretical approaches, the symmetry of the macroscopic order and the corresponding patterns is typically dictated a priori by the assumed microscopic symmetry of the specific active matter model under consideration~\cite{Chate2020} [Fig.\ref{fig:Symmetries_Model}(A)]: models with polar interaction symmetry exhibit polar waves~\cite{Vicsek1995,Chate2004,Chate2008,Bertin2006,Bertin2009}, while models with nematic interaction symmetry show bands (lanes) within which the agents are (preferentially) oriented in parallel~\cite{Chate2006, Ginelli2010,Peshkov2012}.
Thus, in all of these theoretical models, the choice of the underlying microscopic interaction symmetry largely determines the model's phenomenology.
%
%
%
This should be seen in light of the observation that in nature or in the laboratory microscopic details of the agent's propulsion mechanism and interactions are often unclear or essentially inaccessible. Moreover, these properties of the agents might not even be inherent features (traits) characterized by a fixed set of parameters, but could in principle dynamically depend on the emergent collective behavior of the agents, as suggested for animal herds~\cite{Berdahl2018} or chemical active systems~\cite{dauchot2019, Needleman2017}.

Recently, experimental studies of actomyosin motility assays reported coexistence of polar and nematic patterns, with actin filaments dynamically cycling between polar waves and nematic band patterns~\cite{Huber2018}. 
Supported by large-scale computer simulations---emulating the microscopic features of the observed collision statistics---these authors concluded that in this particular case the symmetry of the self-organized patterns is not determined \textit{a priori} by the symmetry of the pairwise interaction between particles, but is itself an \textit{emergent phenomenon} of the many-body system~\cite{Huber2018}. What then is the mechanism underlying this startling phenomenon?

Agent-based simulations indicate that both nematic and polar-ordered clusters can arise when the microscopic alignment between agents is predominantly nematic with a polar contribution, either due to the interactions between extended rods~\cite{Baer2020,Abkenar2013,Wensink14308} or memory in the orientational noise~\cite{Nagai2015}. Furthermore, the existence of distinct regimes of either nematic or polar patterns has been observed in theoretical studies with explicitly mixed alignment symmetries~\cite{Grossmann2015, Ngo2012, Menzel2012}. However, cycling and transformations between patterns of different symmetries as observed in Ref.~\cite{Huber2018} were not reported and the theoretical mechanism behind this coexistence is poorly understood~\cite{Huber2018,Elgeti_2015,Chate2020}. Specifically, neither kinetic nor continuum hydrodynamic approaches have so far been able to reproduce or elucidate this phenomenology~\cite{Elgeti_2015,Chate2020}.

While theoretical approaches with mixed microscopic alignment symmetries have considered alignments that depend on inter-particle distance~\cite{Grossmann2015}, chance~\cite{Ngo2012} or particle species~\cite{Menzel2012}, the computational analysis in Ref.~\cite{Huber2018} suggested that the emergence of dynamic coexistence critically depends on the simultaneous presence of polar and nematic contributions in the binary collision statistics. Here, motivated by the results in Ref.~\cite{Huber2018}, we propose a kinetic theory for a dilute system of propelled particles with tunable `binary collision statistics' [Fig.\ref{fig:Symmetries_Model}(B)]. 
Specifically, we employ a kinetic Boltzmann approach~\cite{Peshkov2014} where particles undergo binary collisions that lead to nematic alignment with a small (tunable) polar bias [Fig.~\ref{fig:Symmetries_Model}(C)].

For both vanishing and fully polar bias, our model recovers the well-studied scenarios of purely nematic~\cite{Peshkov2012} and purely polar~\cite{Bertin2006,Bertin2009} interaction symmetry, respectively. 
Interestingly, for an intermediate polar bias, our model features a transition from macroscopic nematic order at intermediate densities to macroscopic polar order at high densities.
In a regime characterized by intermediate polar bias and intermediate densities, we observe that patterns of polar and nematic symmetry coexist and are dynamically interconvertible, which is reminiscent of the observations in Ref.~\cite{Huber2018}. 
Based on a combination of stability analyses and numerical simulations we argue that such coexistence depends on the inextricable link between symmetry breaking and pattern formation. 
For instance, while the system forms nematic bands in a density regime that leads to symmetry breaking towards macroscopic nematic order, the density at the core of these bands increases and eventually exceeds the threshold value for a transition from macroscopic nematic to macroscopic polar order. 
This spatially local crossing of a critical value in the particle density---a control parameter---triggers local symmetry breaking, which induces the self-organized formation of polar waves.

To substantiate this hypothetical mechanism as a general mechanism for the coexistence of polar and nematic patterns in active-matter systems, we study simplified hydrodynamic equations which capture pattern formation in a nematic phase as well as a transition from macroscopic nematic to polar symmetry for high densities. Indeed, like our kinetic Boltzmann approach, our hydrodynamic theory exhibits a regime of coexisting polar wave and nematic band patterns.    
%
%

Our study thus reveals an interesting \textit{mutual feedback} between pattern formation and macroscopic symmetry breaking in active matter. This feedback occurs because the particle density, which shows pattern formation in active systems, is at the same time a control (bifurcation) parameter for the macroscopic symmetry of the system. 
This twofold role of the particle density transforms symmetry breaking in active systems from a ordering phenomenon under the control of a global parameter into a self-organisation phenomenon with a local interplay between pattern formation and symmetry breaking.
We would argue that this interplay represents a fairly general mechanism that allows macroscopic symmetries to be an emergent property in themselves, rather than being imposed directly by microscopic interaction rules.


\section*{Results}
\subsection*{Kinetic Boltzmann approach with polar and nematic contributions}
\label{Sec:KinBoltzmann}

Our starting point for a mesoscopic theory of aligning agents is the kinetic Boltzmann equation~\cite{Peshkov2014}.
It describes the temporal evolution of the one-particle distribution function $f(\vec{r},\theta,t)$ for the position 
$\vec{r} \,{\in}\, \mathcal{R}^2$ and the orientation $\theta \,{\in}\, [0,2\pi)$ of self-propelled particles that undergo binary aligning collisions in a dilute (dry) system~\cite{Peshkov2014}. It reads 
\begin{align}
\label{Eq::BoltzmannEq}
    \partial_t f(\vec{r},\theta,t) + 
    v_0 \, \vec{e}_\theta^{} \cdot 
    \partial_{\vec{r}} f(\vec{r},\theta,t)
    =
    \mathcal{I}_\text{diff}[f]+\mathcal{I}_\text{coll}[f,f]
    \, ,
\end{align}
where $v_0$ denotes the constant speed of the active particles, $\vec{e}_\theta^{}$ is a unit vector pointing along direction $\theta$, and the terms $\mathcal{I}_\text{diff}[f]$ and $\mathcal{I}_\text{coll}[f,f]$ describe diffusion of individual particles and collisions between particles, respectively (see \textit{SI Appendix}, Note 1).
In more detail, spherical particles (with diameter $d$) are assumed to move ballistically with constant speed $v_0$ along their orientations $\theta$ and can change their orientation by diffusion as well as by local binary collision.
Diffusion is modeled by a shift in a particle's orientation from $\theta$ to $\theta \,{+}\, \eta$ at a rate $\lambda$, where we assume $\eta$ to be a Gaussian-distributed random variable with standard deviation $\sigma$. 
Binary collisions between particles are emulated through `alignment rules' [Fig.~\ref{fig:Symmetries_Model}(B,C)], with an additive random contribution also drawn from a Gaussian distribution with a standard deviation $\sigma'$; in the following we set $\sigma' \,{=}\, \sigma$ for simplicity. 
In the context of the kinetic Boltzmann approach, a fully nematic interaction rule dictates that particles that collide at an acute angle adopt their average orientation (polar alignment), while particles colliding at an obtuse angle also align, but with opposite orientations (anti-polar alignment). For fully polar alignment, particles adopt their average orientation irrespective of their pre-collision angle [Fig.~\ref{fig:Symmetries_Model}]. 

The dynamics of orientational order in the kinetic Boltzmann approach is most conveniently studied by exploiting the fact that the polar vector $\vec{P}$ and nematic tensor $\tensor{Q}$ can be expressed in terms of the Fourier modes $f_k(\vec{r},t) =\int_{-\pi}^{\pi} \!\mathrm{d}\theta \,\e^{\imag \theta k}f(\vec{r},\theta,t)$ of the one-particle distribution function:
\begin{align}\label{Eq::orderdefs}
    \rho\vec{P} 
    = 
    \left(
    \begin{array}{c}
    \Real [f_1]\\ \Imag[ f_1]
    \end{array}
    \right), 
    \quad 
    \rho \tensor{Q}
    =
    \frac{1}{2}
    \left(
    \begin{array}{cc}
    \Real [f_2] & \Imag [f_2] \\ \Imag [f_2] &-\Real[ f_2]
    \end{array} 
    \right).
\end{align} 
Furthermore, the local particle density $\rho(\vec{r},t)$ is given by the $k \,{=}\, 0$ mode: $\rho(\vec{r},t) \,{=}\, f_0(\vec{r},t)$. The dynamics of $f_k(\vec{r},t)$ reads
\begin{align}
    \partial_t f_k 
    &+
    \frac{v_0}{2} 
    \Big[
    \partial_x (f_{k+1}+f_{k-1})-\imag\partial_y (f_{k+1}-f_{k-1})
    \Big]
    \notag \\
    &=
    -\lambda
    \big( 
    1-\e^{-\frac12 k\sigma^2} 
    \big) 
    f_k + 
    \summe{n}{-\infty}{\infty}\mathcal{I}_{n,k}f_n f_{k-n}
    \,.
\label{Eq::BoltzmannFT}
\end{align}
Explicit expressions for the collision coefficients $\mathcal{I}_{n,k}$ can be found in \textit{SI Appendix}, Note 1.
For $k=0$, \eqref{Eq::BoltzmannFT} yields the continuity equation $\partial_t \rho(\vec{r},t)=-v_0\vec{\nabla}\cdot(\rho\vec{P})$.

Solutions of \eqref{Eq::BoltzmannFT} for fully polar~\cite{Bertin2006,Bertin2009} or fully nematic~\cite{Peshkov2012} alignment rules, show a transition from disorder, i.e.\ vanishing polar and nematic order, to nonzero polar or nematic order, respectively, for sufficiently high densities or low noise level $\sigma$. 
Close to the onset of order, it predicts the formation of patterns, consistent with experimental observations and numerical simulations~\cite{Vicsek2012,Wensink14308}. 
The kinetic Boltzmann equation thus serves as a useful basis for a qualitative study of the phenomenology of dilute systems of self-propelled particles.

Recent experimental results from the actin motility assay and corresponding simulation results from agent-based models~\cite{Huber2018} strongly suggest that the relative weights of polar and nematic contributions to the binary collision statistics are critical for the self-organization of spatio-temporal patterns. 
As a minimal extension of fully polar or nematic alignment rules~\cite{Peshkov2014}, we propose a collision rule with a small tunable polar bias. 
Specifically, we assume that colliding particles align in a polar manner when their velocities form an angle difference smaller than $\frac{\pi}{2} \,{+}\, \polarbias$ with $\polarbias \in [0,\frac{\pi}{2}]$ and align anti-polar otherwise [Fig.~\ref{fig:Symmetries_Model}(B,C)]. 
The parameter $\polarbias$ thus characterizes the polar bias, where for $\polarbias \,{=}\, 0$ and $\polarbias \,{=}\, \frac{\pi}{2}$ the collision rule reduces to fully nematic and fully polar collisions, respectively. 
It is convenient to 
rescale time, space, and density such that $v_0 \,{=}\, \lambda \,{=}\, d \,{=}\, 1$. 
Then, the only remaining free parameters are the noise amplitude $\sigma$, the polar bias $\polarbias$, and the mean particle density $\rhohom \,{=}\, \frac1A \int_A\!\mathrm{d}\vec{r} \int_{-\pi}^{\pi}\!\mathrm{d}\theta \, f(\vec{r},\theta,t)$ measured in units of $\lambda/(d v_0)$, i.e.\ the number of particles found within the area traversed by a particle between successive diffusion events. 

\subsection*{Phase diagram and a nonlinear transition to polar order}
\label{Sec:PhaseDiagram}

\begin{figure}[t]
\centering
\includegraphics[width=1.\linewidth]{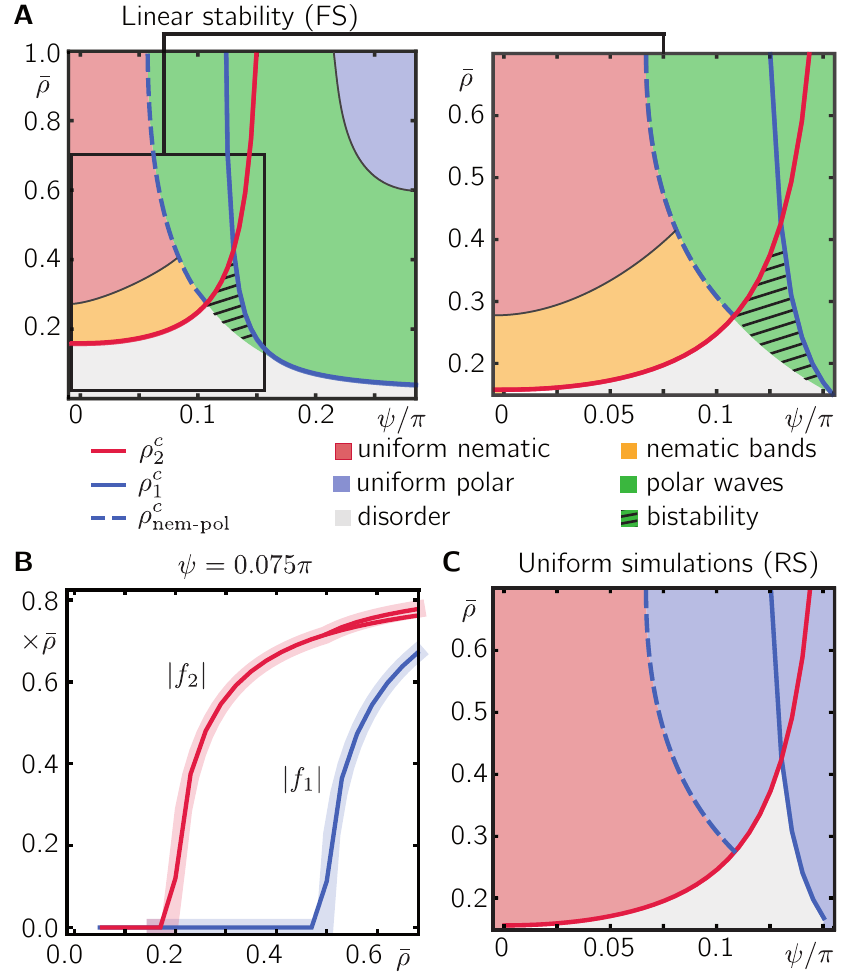}
\caption{\textbf{Uniform solutions and linear stability.} \textbf{(A)} Regimes and linear stability of spatially uniform solutions for  \eqref{Eq::BoltzmannFT} in angular Fourier space (FS) with a truncation at $k_c \,{=}\, 10$. Squares denote stable solutions, triangles indicate solutions that are unstable against spatial perturbations, suggesting pattern formation. Solid lines denote the critical transition densities $\rho_2^c$ and $\rho_1^c$ from a disordered solution to nematic and polar order, respectively. Our analysis reveals another transition from nematic order to polar order at intermediate polar bias (blue dashed line serves as guide to the eye). \textbf{(B)} Spatially uniform solutions for $f_1$ and $f_2$ at $\polarbias \,{=}\, 0.075\pi$ calculated from the truncated Boltzmann equation in angular Fourier space (solid lines) as compared to spatially uniform solutions of the Boltzmann equation in real space (RS,~\eqref{Eq::BoltzmannEq}, shaded lines) calculated using the generalized \texttt{SNAKE} algorithm. \textbf{(C)} Phase portrait of spatially uniform solutions using the generalized \texttt{SNAKE} algorithm. Simulations were done on a single lattice point starting at a disordered state with small fluctuations in the angular distribution.  The noise value was fixed to $\sigma \,{=}\, 0.2$.}
\label{fig:LinStabPhaseDiagram}
\end{figure}

Since the collision coefficients $\mathcal{I}_{n,k}$ are zero for $k \,{=}\, 0$, \eqref{Eq::BoltzmannFT} possesses a spatially uniform solution with zero order, i.e. $f_k \,{=}\, 0$ for $|k| \,{>}\, 0$, and uniform density $f_0 \,{=}\, \rhohom$.
Up to linear order, a small perturbation $\delta f_k$ of this disordered state evolves according to $\partial_t \delta f_k(t) \,{=}\, \mu_k(\rhohom,\sigma,\polarbias) \, \delta f_k$ with the growth rate $\mu_k(\rhohom,\sigma,\polarbias) \,{=}\, (\mathcal{I}_{0,k} \,{+}\, \mathcal{I}_{k,k})\rhohom \,{-}\, \lambda (1 \,{-}\, \e^{-\frac12 k\sigma^2})$. 
The zeros of these growth rates, $\mu_k(\rho^c_k,\sigma,\polarbias) \,{=}\, 0$, mark the critical densities $\rho^c_k(\sigma,\polarbias)$ above which the mode $k$ grows exponentially. 
While previous studies have focused on the onset of order for fully polar and nematic interactions as a function of the density $\rhohom$ and noise amplitude $\sigma$~\cite{Peshkov2014}, in the following we keep the noise level constant, $\sigma \,{=}\, 0.2$, and focus on the onset of order as a function of the polar bias $\polarbias$.
Figure~\ref{fig:LinStabPhaseDiagram}(A) shows the critical densities $\rho^c_1(\polarbias)$ and $\rho^c_2(\polarbias)$ for the onset of polar and nematic order, respectively. 
For small and large polar bias, only the growth rate for $k\,{=}\,2$ or $k\,{=}\,1$, respectively, changes sign, 
indicating that there are transitions from a disordered state to a state with either nematic order (for small polar bias $\psi$) or polar order (for large polar bias $\psi$).
In contrast, for intermediate polar bias, the transition densities $\rho^c_1(\polarbias)$ and $\rho^c_2(\polarbias)$ cross, implying that there is a regime in the ($\rhohom$, $\psi$) phase diagram where the disordered state is linearly unstable under both polar and nematic perturbations. 

After identifying the parameter regimes where the spatially uniform disordered solutions become unstable, we now determine the stable, spatially uniform, ordered solutions of~\eqref{Eq::BoltzmannFT} in these regimes.
As this is no longer feasible analytically (due to the infinite sum), we resort to approximate solutions, exploiting the fact that even above the ordering transition, modes with sufficiently large $|k|$ are still negligible~\cite{Bertin2009,Denk2016}.
To this end, following Ref.~\cite{Denk2016}, we set all spatial derivatives and all Fourier modes $f_k$ beyond a certain cutoff $k_c$ in~\eqref{Eq::BoltzmannFT} to zero, and numerically solve the ensuing equations for all remaining modes with $|k| \,{\leq}\, k_c$. 
We then performed a linear stability analysis of the resulting spatially uniform solutions against uniform as well as nonuniform (wave-like) perturbations. 
The directions of spatial perturbations were varied to probe for instabilities perpendicular and parallel to the orientation of the spatially uniform (polar and nematic) order parameters. 
Based on this analysis we identified the type of order exhibited by spatially uniform solutions, as well as their stability against wave-like perturbation for different values of the average density $\rhohom$ and polar bias $\polarbias$ [see Fig.~\ref{fig:LinStabPhaseDiagram}(A)].

Above the critical transition densities $\rho^c_2(\polarbias)$ and $\rho^c_1(\polarbias)$, we indeed find spatially uniform solutions with nonzero nematic and polar order, respectively, as suggested by the linear stability analysis of the disordered state.
In addition, within these regimes, we identify subregimes in which the respective spatially uniform solutions are unstable under spatial perturbations, suggesting the formation of spatially \textit{nonuniform} patterns [indicated by the triangles in Fig.~\ref{fig:LinStabPhaseDiagram}(A)]. 
We find that slightly above the transition density to nematic order, $\rho_2^c$, spatially uniform nematic order is unstable against wave-like perturbations perpendicular to the nematic order, which suggests formation of \textit{nematic band patterns}.
Furthermore, in a subregime of polar order, uniform polar order is unstable against spatial perturbations parallel to the orientation of polar order, which suggests the emergence of travelling \textit{polar waves}.
The prediction of nematic bands and polar waves for small and large polar bias is in accordance with previous studies on systems with either fully nematic or polar interaction symmetries, respectively~\cite{Peshkov2014}.
Interestingly, for sufficiently large polar bias ($\polarbias \,{\gtrsim}\, 0.05\pi$) and high enough densities ($\rhohom \,{>}\, \rho_2^c$) we identify a transition that has not been observed in previous studies with fully nematic or polar alignment: as the density is increased above $\rhonempol(\polarbias)$ [indicated by the dashed line in Fig.~\ref{fig:LinStabPhaseDiagram}(A)], there is a direct transition from solutions with uniform nematic order to solutions with uniform polar order. 
Furthermore, within the regime of linearly stable disorder, and close to the intersection of $\rho_1^c$ and $\rho_2^c$, our analysis reveals a regime of \textit{bistability}. 
Here, we find both linearly stable disorder and uniform polar order, which is linearly unstable to polar wave patterns. 
This bistability of disorder and polar patterns is an interesting topic in its own right and is in fact closely related to the observation of a discontinuous transition between the isotropic and the ordered phase in the homogeneous regime as indicated for aligning hard discs~\cite{Lam_2015,Lam_2015_2}. We will defer the study of this bistability regime to future work and focus here on the transition from nematic to polar order. 

To independently test the predictions of our approximate solutions and linear stability analyses, we numerically solved the Boltzmann equation,~\eqref{Eq::BoltzmannEq}, in real space (for details see \textit{SI Appendix}, Note 2). 
To this end, we used a modified version of the \texttt{SNAKE} algorithm~\cite{Flo2014}. To determine the uniform solutions, we simulated a spatially uniform system [see Fig.~\ref{fig:LinStabPhaseDiagram}(C)] and find excellent agreement with the approximate solutions of our spectral analysis [Fig.~\ref{fig:LinStabPhaseDiagram}(B), compare (A) with (C)].

\subsection*{Pattern formation leads to dynamic transformations between nematic and polar symmetries}
\label{Sec:patternsandtransition}

\begin{figure}[!b]
\centering
\includegraphics[width=\linewidth]{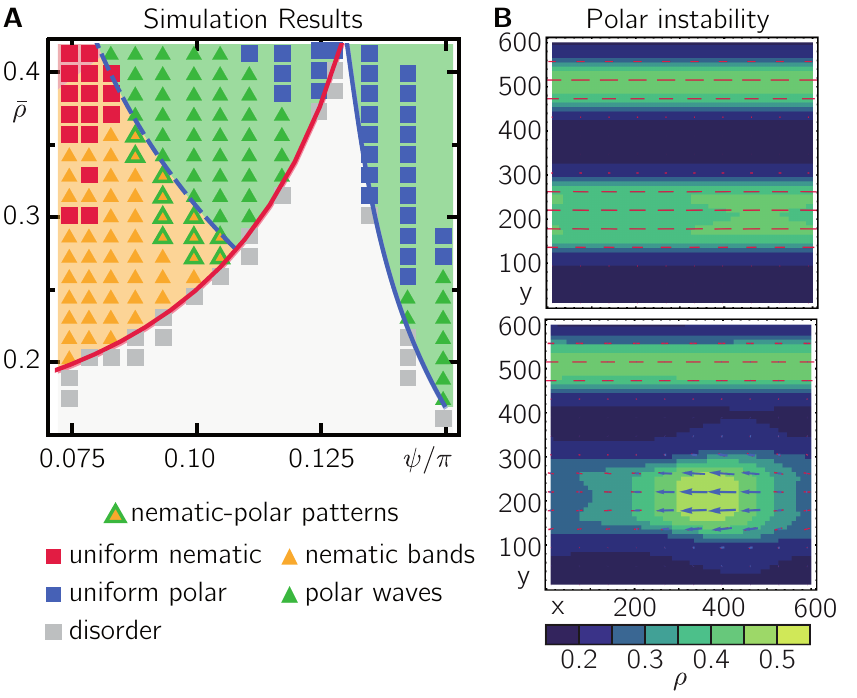}
\caption{\textbf{Dynamic transformations between nematic and polar patterns.} \textbf{(A)} Numerical solutions of~\eqref{Eq::BoltzmannEq} display regimes of uniform nematic and polar order as well as nematic band and traveling wave patterns (symbols as in Fig.~\ref{fig:LinStabPhaseDiagram}(A)). The green and orange shades denote regimes in which linear stability analysis predicts polar and nematic patterns, respectively (regimes in in Fig.~\ref{fig:LinStabPhaseDiagram}(A,C)). In addition to the states predicted by linear stability analysis (Fig.~\ref{fig:LinStabPhaseDiagram}), we find dynamic transformations between patterns of nematic to polar patterns below $\rhonempol$. \textbf{(B)} Simulation snapshots in the regime of dynamic transformations (nematic-polar patterns) shortly before (top) and after (bottom) a local instability within the core of a nematic band. Red bars and blue arrows indicate the orientation and strength of local nematic and polar order, respectively. All simulations started from uniform disordered states with small random fluctuations. For details on the simulations see \textit{SI Appendix}, Note 2.}.
\label{fig:Transitions}
\end{figure}

To resolve the spatio-temporal dynamics of the kinetic Boltzmann equation, \eqref{Eq::BoltzmannEq}, especially in regimes where our stability analysis predicts spatially uniform solutions to be unstable, we employed our modified \texttt{SNAKE} algorithm for a spatially extended system to numerically solve the Boltzmann equation~\eqref{Eq::BoltzmannEq} in extensive parameter sweeps in both the global density and polar bias. [Fig.~\ref{fig:Transitions}(A)] shows simulation results (symbols) against the background of our predictions from linear stability analysis (shaded) [Fig.~\ref{fig:LinStabPhaseDiagram}(A)]. 

For vanishing and small polar bias, our numerical simulations show nematic band solutions for densities slightly above $\rho^c_2(\polarbias)$ and uniform nematic states at higher densities [orange and red symbols in Fig.~\ref{fig:Transitions}(A), respectively; see \textit{SI Appendix}, Note 2, Fig.S1(A)].
For larger polar bias, we find regimes of traveling-wave solutions and spatially uniform polar-ordered states consistent with our predictions from linear stability analysis [green and blue symbols in Fig.~\ref{fig:Transitions}(A), respectively; see \textit{SI Appendix}, Note 2, Fig.S1(B)].
The regimes of patterns and uniformly ordered solutions are in good agreement with our linear stability analysis [Fig.~\ref{fig:LinStabPhaseDiagram}(A)].

Remarkably, we find patterns with polar order even for densities below $\rhonempol$  [nematic-polar patterns in  Fig.~\ref{fig:Transitions}(A)]. 
Here, the spatio-temporal dynamics of the system exhibits both patterns with polar and nematic order that dynamically interconvert into each other [Fig.~\ref{fig:Transitions}(B); see \textit{SI Appendix}, Note 2, Movie 1-3]. 
Starting from a disordered state, the nematic order grows quickly and the system starts to exhibit high-density nematic band patterns (in line with linear stability analysis). 
While the ensuing average nematic order is approximately the same as that found for spatially uniform solutions of~\eqref{Eq::BoltzmannFT}, the local nematic order is much higher \textit{within} the nematic bands and approaches zero in the disordered regions between the bands~(see \textit{SI Appendix}, Note 2, Fig.S2). 
Actually, the local density in the center of a band is so high that it by far exceeds the density at the onset of the nematic-polar order transition $\rhonempol$. 
This suggests that within a band, purely nematic order eventually becomes unstable and polar order starts to emerge there. 
Indeed, we observe that after some time, polar order locally builds up in the nematic bands and subsequently leads to the formation of polar waves that propagate along the nematic band [Fig.~\ref{fig:Transitions}(B), \textit{SI Appendix}, Movies 1-3]. 
Initialising the system with different randomly disordered states, we find that this local polar instability leads to various distinct types of spatio-temporal dynamics including polar waves within nematic bands [Fig.~\ref{fig:Transitions}(B), \textit{SI Appendix}, Movie 1], complete replacement of nematic bands by polar waves (\textit{SI Appendix}), Movie 2) and dynamic switching between nematic bands and polar waves (\textit{SI Appendix}, Movie 3). 

In summary, for low and high polar bias our kinetic Boltzmann approach is consistent with the classical conception of self-propelled particle systems with predominantly nematic or polar symmetry~\cite{Chate2020} including the formation of nematic band patterns and traveling polar waves at the onset of nematic and polar order, respectively. 
For moderate polar bias, stability analysis uncovers an additional transition from nematic solutions to polar solutions at a density $\rhonempol \,{>}\, \rho_2^c$. 
This transition gives rise to novel spatio-temporal dynamics that are not predicted by linear stability analysis: 
Our numerical simulations reveal that the high density core of nematic bands can locally cross the threshold density $\rhonempol$, which favors the formation of polar waves. 
This instability eventually results in traveling-wave solutions, as well as more complex dynamics such as coexisting polar waves and nematic bands, and dynamic rearrangements of nematic bands. Importantly, this implies that the symmetry of the emerging patterns critically depends on the dynamics of the system and is not already determined or apparent from the assumed alignment symmetry nor indicated by linear stability analysis. 
This shows that the symmetry of the spatio-temporal pattern is itself an emergent property, whose dynamics is based on a mutual feedback between pattern formation and \textit{local} symmetry breaking due to the redistribution (accumulation) of mass (particle density).

\subsection*{Simple hydrodynamic equations account for coexisting symmetries}
\label{Sec:Hydro}

The kinetic Boltzmann equation serves as a useful starting point for the derivation of hydrodynamic theories for the dynamics close to the onset of order~\cite{Peshkov2014}. 
These field theories, which are more amenable to theoretical analysis than the kinetic Boltzmann equation, have been successfully employed to reproduce and understand the observed phenomenology of active matter systems close to ordering transitions~\cite{Ramaswamy2017,Chate2020}. 

In order to derive a closed set of hydrodynamic equations from the kinetic Boltzmann equation~\eqref{Eq::BoltzmannFT}, previous studies~\cite{Peshkov2014} assumed that, close to the onset of polar or nematic order, the respective order fields $f_1$ and $f_2$ and their temporal and spatial variations, are small. 
This assumption yields scaling relations for $f_k$ which suggest systematic truncation schemes for the infinite sum in~\eqref{Eq::BoltzmannFT} to arrive at closed equations for the dominant hydrodynamic fields. Balancing terms in the Boltzmann equation, Peshkov et al.~\cite{Peshkov2012} have proposed the following scaling relations for a system of polar particles with a fully nematic collision rule ($\polarbias \,{=}\, 0$): $\rho \,{-}\,\rhohom \,{\sim}\, \varepsilon$, $\{f_{2k-1},f_{2k}\}_{k\geq 1} \,{\sim}\, \varepsilon^k$, $\partial_t \,{\sim}\, \varepsilon$ and $\partial_{x/y} \,{\sim}\, \varepsilon$. 
With these scaling relations, one can expand the sum in~\eqref{Eq::BoltzmannFT} retaining only terms up to order $\varepsilon^3$ to get closed equations for the order fields $f_{\{1,2,3,4\}}$. 
The equations for $f_3$ and $f_4$ yield expressions for $f_3$ and $f_4$ in terms of $f_1$ and $f_2$. In addition to the continuity equation for $f_0$, one arrives at the following hydrodynamic equations for $f_1$ and $f_2$:
\begin{subequations}
\label{Eq::Hydroglg}
\begin{align}
    \partial_t f_1
    =&-(\alpha_0+\rho\alpha_1)f_1+\alpha_2 f_1^* f_2-\alpha_3 |f_2|^2f_1\notag\\
    &-\frac{1}{2}(\nabla\rho+\nabla^*f_2)+\gamma_1 f_2^*\nabla f_2 \, ,
    \label{Eq::Hglgf1}\\
    \partial_t f_2
    =&(-\beta_0+\rho \beta_1)f_2+\beta_2 f_1^2-\beta_3|f_2|^2f_2 -\beta'_3|f_1|^2f_2\notag\\
    &{-}\frac{1}{2}\nabla f_1+\gamma_2\nabla\nabla^* f_2-\gamma_3 f_1^*\nabla f_2-\gamma_4\nabla^*(f_1 f_2) 
    \, ,
    \label{Eq::Hglgf2}
\end{align}
\end{subequations}
which are related to the polar and nematic order parameter through~\eqref{Eq::orderdefs}. To simplify the notation, we defined $\nabla \,{:=}\, \partial_x \,{+}\, \imag\partial_y$ and the asterisks denote complex conjugation. 
The coefficients contain contributions from angular diffusion and combinations of the collision integrals $\mathcal{I}_{n,k}$ introduced in~\eqref{Eq::BoltzmannFT} (for explicit expressions see \textit{SI Appendix}, Note 3). 
The first terms in~\eqref{Eq::Hglgf1} and~\eqref{Eq::Hglgf2}, which are linear in the modes $f_1$ and $f_2$, suggest that nematic and polar order grow beyond the densities given by $\rho^c_2 \,{=}\, \beta_0/\beta_1$ and $\rho^c_1 \,{=}\, {-}\alpha_0/\alpha_1$, respectively (these expressions for $\rho^c_1$ and $\rho^c_2$ are in fact equivalent to the expressions discussed above in the linear stability analysis of~\eqref{Eq::BoltzmannFT}). 
In addition, even for densities below $\rho^c_1$, i.e. $(\alpha_0 \,{+}\, \alpha_1\rho) \,{<}\, 0$, the second term in~\eqref{Eq::Hglgf1} might still lead to growth of polar order when $|f_2|$ is large enough and the second term in~\eqref{Eq::Hglgf1} dominates the first. 

Given the validity of~\eqref{Eq::Hydroglg} for systems with a fully nematic collision rule~\cite{Peshkov2012}, one could hope that~\eqref{Eq::Hydroglg} provides a suitable hydrodynamic description in the presence of a (at least small) polar bias. However, when we extend the collision integrals $\mathcal{I}_{n,k}$ and thereby the coefficients in~\eqref{Eq::Hydroglg} towards a polar bias as depicted in~\ref{fig:Symmetries_Model}(B),(C) and calculate the uniform solutions of~\eqref{Eq::Hydroglg} as well as their linear stability, we find that the resulting phase diagram of~\eqref{Eq::Hydroglg} [see \textit{SI Appendix}, Note 3, Fig.S3] critically differs from the phase diagram of~\eqref{Eq::BoltzmannEq} [Fig.~\ref{fig:LinStabPhaseDiagram}(A)]. 
Most importantly, \eqref{Eq::Hydroglg} does not feature a secondary transition from nematically ordered solutions to polar-ordered solutions for moderate polar bias as observed in Fig.~\ref{fig:LinStabPhaseDiagram}(A) (dashed line). 
The discrepancy between the phase diagrams obtained from the solutions of~\eqref{Eq::Hydroglg} and~\eqref{Eq::BoltzmannFT} suggests that in the presence of a polar bias, modes $f_k$ with $|k|>4$, which were neglected in the derivation of~\eqref{Eq::Hydroglg}, become important. 
To test the relevance of modes with $|k|>4$ we performed numeric simulations of the dynamics of the Fourier modes,~\eqref{Eq::BoltzmannFT}, taking into account modes $f_k$ with $|k|$ up to a certain $k_h$ and setting all modes with $|k|>k_h$ and their derivatives to zero (for the numerical solution we used \texttt{XMDS2}~\cite{DENNIS2013201}, a fast Fourier transform (FFT)-based spectral solver).
Consistent with our simulations in real space [Fig.\ref{fig:Transitions}(A)] we find stable nematic bands, nematic bands with polar instabilities, and polar traveling waves for small, moderate, and large polar bias, respectively. 
However, soon after the polar instabilities within nematic bands have triggered polar waves [see Fig.\ref{fig:Transitions}(C)] our simulations show that the ensuing polar order diverges.  
This  divergence in the numerical solution of \eqref{Eq::BoltzmannFT} occurs even for relatively large cutoff mode numbers $k_h \,{\gtrsim}\, 12$ [\textit{SI Appendix}, Note.3A, Fig.S4]. Taken together, these findings suggest that a crude truncation of~\eqref{Eq::BoltzmannFT}, and, in particular, the reduced set of equations for the lowest two modes \eqref{Eq::Hydroglg} with the coefficients determined by a scaling analysis of the kinetic Boltzmann equation, do not (fully) capture the dynamic transition of patterns as observed in Fig.~\ref{fig:Transitions}.


Alternative coarse-grained approaches to active matter have constructed hydrodynamic equations mainly on the basis of symmetry arguments and small amplitude expansions in the order fields and their gradients~\cite{TonerTu1995, Ramaswamy2010, Marchetti2013}. Those approaches study the phenomenology of the hydrodynamic equations as a function of the coupling coefficients, which are considered to be free phenomenological parameters.
In the following, we take a \textit{semi-phenomenological} approach which retains the functional dependencies of~\eqref{Eq::Hydroglg} arrived at by using the Boltzmann equation while exploring the phenomenology of this field theory for general coupling parameters. 
In the spirit of a generalized Ginzburg-Landau approach, we require that the now phenomenological parameters $\beta_0$ and $\beta_1$ are positive in order to reproduce a bifurcation from a disordered state to a nematic state at a critical density $\rho^c_2 \,{=}\, \beta_0/\beta_1$. 
Likewise, to suppress a direct bifurcation from disorder to polar order, $\alpha_0$ and $\alpha_1$ are chosen to be positive as well. 
Under these conditions, uniform polar order only grows when the second term in~\eqref{Eq::Hglgf1} is large enough.
To ensure saturation for $f_1$ and $f_2$, the coefficients for the highest order terms, i.e.\ $\alpha_3$, $\beta_3$, and $\beta'_3$, are chosen to be positive. 
We argue that an increasing polar bias, which mingles polar and nematic collisions, leads to an enhanced coupling strength between fields of nematic symmetry (i.e.\ $f_{2k},\,k \,{\geq}\, 1$) and polar symmetry (i.e.\ $f_{2k-1},\,k \,{\geq}\, 1$). In~\eqref{Eq::Hglgf1}, this \textit{polar-nematic coupling strength} is set by $\alpha_2$.
To systematically study the effect of a varying polar-nematic coupling strength $\alpha_2$, we fixed all coefficients in~\eqref{Eq::Hydroglg} except for $\alpha_2$. 
For specificity, all other coefficients were fixed to  values derived from the kinetic Boltzmann equation with fully nematic alignment and slightly above the transition to nematic order ($\rho=\rhohom \,{=}\, 0.16$ and $\sigma \,{=}\, 0.2$). This choice naturally satisfies all the above-mentioned general conditions on the coefficients. 

\begin{figure}[t]
\centering
\includegraphics[width=1.\linewidth]{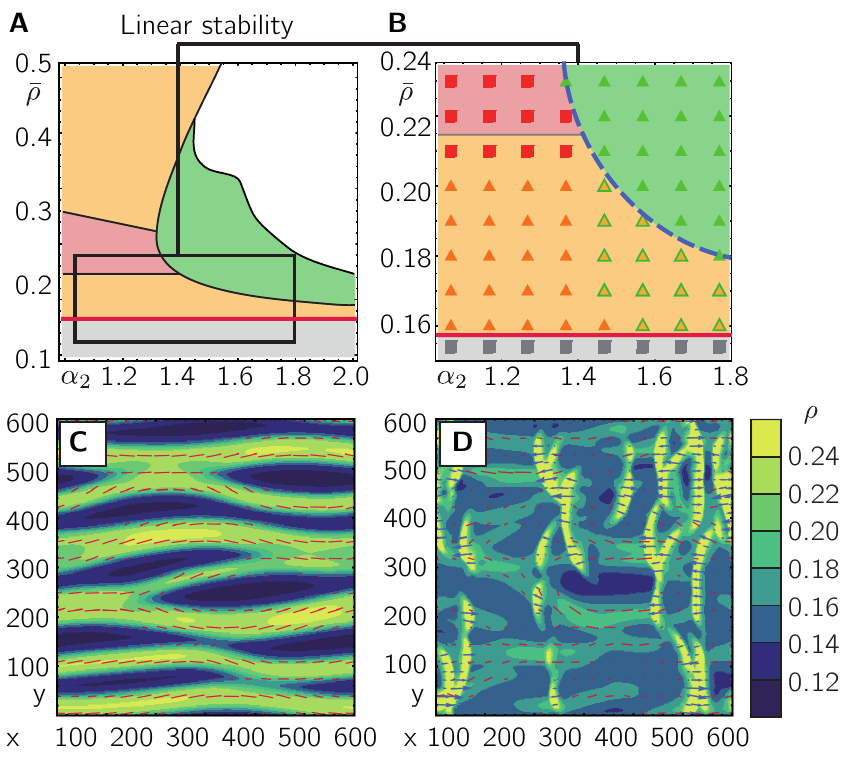}
\caption{
\textbf{Cycling symmetries in hydrodynamic approach.} \textbf{(A)} Linear stability analyses of~\eqref{Eq::Hydroglg} yields a phase diagram with regimes of polar and nematic patterns that qualitatively resembles the phase diagram derived from the original kinetic Boltzmann approach for a mixed collision rule [Fig.~\ref{fig:LinStabPhaseDiagram}; colors denote regimes of different symmetries and patterns as in Fig.~\ref{fig:LinStabPhaseDiagram}(A) and ~\ref{fig:Transitions}(A)].  \textbf{(B)} Simulations [denoted by symbols, as in Fig.~\ref{fig:Transitions}(A)] show the formation of nematic bands and polar waves, as well as coexistence patterns with transformations between nematic and polar patterns. \textbf{(C)} Snapshot of nematic band patterns shortly before the onset of local polar instabilities (red bars indicate orientation and strength of local nematic order). \textbf{(D)} At later times, polar instabilities lead to the formation of traveling wave patterns with complex dynamics including coexistence of local nematic and polar ordered regions which interact and transform into each other. Blue arrows denote the strength and direction of polar order. (Parameters are $\alpha_2 \,{=}\, 1.5$, $\rhohom \,{=}\, 0.18$.). For details on the simulations see \textit{SI Appendix}, Note 3B}
\label{fig:HydroApproach}
\end{figure}

By calculating the spatially uniform solutions of~\eqref{Eq::Hydroglg} and their linear stability against uniform and nonuniform perturbations, we obtain the phase diagram as a function of the average density $\rhohom$ and the polar-nematic coupling strength $\alpha_2$ [see Fig.~\ref{fig:HydroApproach}(A)]. 
This phase diagram shares key qualitative similarities with our linear stability analysis of the kinetic Boltzmann equation [Fig.~\ref{fig:LinStabPhaseDiagram}(A)]. By construction, for average densities above $\rho^c_2$ the disordered solution is unstable and we find uniform solutions with nematic order [orange and red areas in Fig.~\ref{fig:HydroApproach}(A)].  Slightly above $\rho^c_2$, these solutions are stable against uniform perturbations, but unstable against nonuniform perturbations perpendicular to the orientation of nematic order, suggesting the formation of nematic band patterns [orange area in Fig.~\ref{fig:HydroApproach}(A)]. For moderate coupling strengths $\alpha_2$ ($\alpha_2\gtrsim 1.4$ in units $\left[\text{density}/\text{time}\right]$), our linear stability analysis suggests a direct transition at $\rhonempolhydro$ from a phase of nematic band patterns to a phase of polar patterns [dotted line in Fig.~\ref{fig:HydroApproach}(A)], similar to that observed in our analysis of the kinetic Boltzmann equation [dashed line in Fig.~\ref{fig:LinStabPhaseDiagram}(A)].
When both $\alpha_2$ and $\rhohom$ are large, there are no physical solutions (the polar order lies beyond the attainable density), indicating that our hydrodynamics equations are not adequate in these regimes. At small $\alpha_2$ and large $\rhohom$ we find re-entrance into a regime of nematic patterns. In the following we will disregard these regimes and focus on the regime close to the transition density $\rho_2^c$ [see highlighted regime in Fig.~\ref{fig:HydroApproach}(A)].

To resolve the spatio-temporal dynamics of the modes $f_1$ and $f_2$ beyond linear stability analyses, we numerically solved equations~\eqref{Eq::Hydroglg}, together with the continuity equation, using \texttt{XMDS2} software. 
For low $\alpha_2\lesssim1.3$ and slightly above $\rho_c^2$, we find nematic band patterns as predicted by our linear analysis. 
For larger $\alpha_2$ and densities between $\rho_2^c$ and $\rhonempolhydro$, the system shows dynamic transitions between patterns of polar and nematic symmetry [see Fig.~\ref{fig:HydroApproach}(B); \textit{SI Appendix}, Movie 4] reminiscent of the observations in our kinetic Boltzmann approach [Fig.~\ref{fig:Transitions}](B)]. First, the system forms nematic bands as predicted by linear stability analysis [Fig.~\ref{fig:HydroApproach}(A)]. At the core of these bands the local density exceeds $\rhonempolhydro$, suggesting a local instability towards polar order (\textit{SI Appendix}, Note 3B, Fig.S5). Accordingly, we observe the formation of travelling wave patterns along the nematic bands. These instabilities eventually result in intriguing spatio-temporal patterns of nematic and polar order which dynamically interconvert in a cyclic fashion [Fig.~\ref{fig:HydroApproach}(C); \textit{SI Appendix}, Note 3B, Movie 4].

\section*{Discussion}

Motivated by the intriguing dynamic coexistence of polar and nematic patterns observed in recent active matter experiments and simulations~\cite{Huber2018}, we studied a system of self-propelled particles that exhibit binary nematic alignment interactions with a tunable polar contribution. For a moderate polar bias, our kinetic Boltzmann approach reveals a direct transition from a phase of macroscopic nematic to polar order for high enough densities. In addition to the previously studied nematic bands and traveling waves for respectively small and large polar bias~\cite{Chate2020,Peshkov2014}, we identify a parameter regime of moderate polar bias and density where a density increase in nematic patterns can induce a local symmetry-breaking instability and an ensuing transition to polar patterns. Depending on system size and initial conditions, this dynamic transition can lead to different final patterns, including coexistence of patterns with nematic and polar symmetry as well as dynamic transformations between them. We find a similar phenomenology in hydrodynamic equations when the coupling between polar and nematic order is strong enough. 

Our findings shed new light on traditional symmetry assumptions in dilute active-matter theories~\cite{Chate2020} and suggest that the symmetry of patterns can depend on the (nonlinear) dynamics of the system. In the system we studied, a \textit{global} symmetry-breaking instability of the uniform nematic state first leads to a redistribution of the density into nematic band patterns. Since in our system the density acts as control (bifurcation) parameter for the macroscopic symmetry, the high-density core of a nematic band can locally cross a threshold value in the density such that there is symmetry breaking, i.e.\ the symmetry of the system changes from nematic to polar. These \textit{local} symmetry-breaking transitions eventually lead to the self-organized coexistence of, and cycling between polar and nematic patterns. In contrast to coexistence of spatially distant patterns~\cite{Baer2020}, here nematic and polar patterns are thus firmly intertwined: nematic bands serve as scaffolds for the creation of polar wave patterns, which propagate along the nematic bands and decay in its low-density neighborhood, which again fuels the formation of nematic bands. All of these observations are in very good agreement with the phenomenology observed in previous actomyosin motility assays and agent-based simulations of rods~\cite{Huber2018}.

Previous studies have observed instabilities of nematic band structures in systems with fully nematic alignment interactions between polar particles~\cite{Giomi_2012, Wensink14308}, particles with velocity reversal~\cite{Grossmann2016} and apolar particles~\cite{Ngo2014, Cai2019}. There, for large enough system sizes, nematic bands exhibit a \textit{transverse} instability, which causes long undulations and transverse break-up of nematic bands and can lead to chaotic dynamics~\cite{Giomi_2012, Wensink14308, Ngo2014, Cai2019}. While our large-scale simulations of the hydrodynamics equations also exhibit undulations of nematic bands [Fig.~\ref{fig:HydroApproach}(B,C); \textit{SI Appendix}, Movie 4], our system with mixed alignment symmetry features an additional instability of nematic bands towards polar order \textit{parallel} to nematic bands, which leads to the formation of polar waves along the bands. 

Our study highlights a \textit{mutual feedback} between pattern formation and local symmetry-breaking instabilities (bifurcations) as the cause of dynamic coexistence between patterns of different symmetry. In a different but related context, the role of local equilibria and their stability has been studied for mass-conserving reaction-diffusion systems~\cite{Halatek2018,brauns2018phasespace}.
We hypothesize that this feedback is not limited to our study, but could be a more general principle whenever a control parameter (such as density) is dynamically redistributed during pattern formation. From a broader perspective on biological active matter, this could apply whenever individuals dynamically change their microscopic properties (velocity, interaction behavior, etc.) in response to macroscopic parameters such as the density. Prominent examples of such feedback between macroscopic effects and the microscopic components of the system are found synthetic active systems with chemical interactions~\cite{dauchot2019}, collective sensing in bacteria~\cite{Hammer2003,Rein2016,Camley_2018}, and animals~\cite{Berdahl2018}.

\section*{Bibliography}

\onecolumn

\newpage

\captionsetup*{format=largeformat}

\section{Kinetic Boltzmann equation}

Following Refs.~\cite{Bertin2006,Bertin2009}, the kinetic Boltzmann equation for the orientational one-particle distribution function $f(\vec{r},\theta,t)$ reads:
\begin{align}
\label{Eq::SUPBoltzmannEq}
\partial_t f(\vec{r},\theta,t)+v_0 \vec{e}(\theta)\cdot\partial_\vec{r}f(\vec{r},\theta,t)=\mathcal{I}_\text{diff}[f]+\mathcal{I}_\text{coll}[f,f]\,,
\end{align}
where $\mathcal{I}_\text{diff}$ and $\mathcal{I}_\text{coll}$ denote the diffusion and collision integrals, respectively. They are given by
\begin{subequations}
\begin{align}
\mathcal{I}_d[f]=&-\lambda f(\theta)+\lambda\integ{-\pi}{\pi}{\theta'}\integ{-\infty}{\infty}{\eta}f(\theta')\,P_\sigma(\eta)\delta_{2\pi}(\theta'-\theta+\eta)\,, \\
\mathcal{I}_c[f]=& -f(\theta)\integ{-\pi}{\pi}{\theta'}\, \mathcal{R}(\theta,\theta')\,f(\theta')+\integ{-\pi}{\pi}{\theta_1}f(\theta_1)\integ{-\pi}{\pi}{\theta_2}
\, \mathcal{R}(\theta_1,\theta_2)f(\theta_2)\,\integ{-\infty}{\infty}{\eta} P_\sigma(\eta)\,\Psi_\eta(\theta_1,\theta_2,\theta)\,.
\end{align}
\end{subequations}
$P_\sigma(\eta)$ is a Gaussian distribution with standard variation $\sigma$ and $\delta_{2\pi}$ denotes a generalized Kronecker delta, imposing that the argument is zero modulo $2\pi$. $\mathcal{R}(\theta_1,\theta_2)$ denotes the differential cross section of two particles with orientations $\theta_1$ and $\theta_2$. For disc-like particles with diameter $d$, $\mathcal{R}(\theta_1,\theta_2)$ is given by $\mathcal{R}(\theta_1,\theta_2)=4v_0 d\sin(\frac{\theta_1-\theta_2}{2})$~\cite{Bertin2006,Bertin2009}. The binary interaction rule enters through the alignment function $\Psi_\eta (\theta_1,\theta_2,\theta)$:
\begin{align*}
\begin{array}{ll}
\text{For polar alignment: }&     \Psi_\eta(\theta_1,\theta_2,\theta)
=\delta_{2\pi} \left(\frac{\theta_1-\theta_2}{2}-\theta+\eta \right) \\
 & \\
\text{For antipolar alignment: }& \Psi_\eta(\theta_1,\theta_2,\theta)=\frac{1}{2}\delta_{2\pi}\left(\frac{\theta_1-\theta_2}{2}-\theta+\frac{\pi}{2}+\eta\right) +\frac{1}{2}\delta_{2\pi}\left(\frac{\theta_1-\theta_2}{2}-\theta-\frac{\pi}{2}+\eta\right)
\end{array}
\end{align*}
For an interaction rule with variable polar bias $\polarbias$ we assume polar alignment for an intermediate angle with $|(\theta_1-\theta_2)|<\pi/2+\polarbias$ and antipolar alignment otherwise. The parameter $\polarbias\,\epsilon\,[0,\pi/2]$ thus characterizes the strength of the polar bias where for $\polarbias=0$ and $\polarbias=\pi/2$ the collision rule reduces to fully nematic or polar collisions, respectively. 

In the following, we rescale time, space, and density such that $v_0=\lambda=d=1$. Then, the only remaining free parameters are the noise amplitude $\sigma$, the polar bias $\polarbias$, and the mean particle density $\bar{\rho}=A^-1\int_A\!\mathrm{d}\vec{r}\int_{-\pi}^{\pi}\!\mathrm{d}\theta\,f(\vec{r},\theta,t)$ measured in units of $\lambda/(d v_0)$, i.e., the number of particles found within the area traversed by a particle between successive diffusion events. In order to study solutions of the the kinetic Boltzmann equation \eqref{Eq::SUPBoltzmannEq}, it is convenient to expand this equation for $f$ in terms of Fourier modes of the angular variable given by
\begin{align}
\label{angularFT}
f_k(\vec{r},t) =\integ{-\pi}{\pi}{\theta}\,\e^{\imag \theta k}f(\vec{r},\theta,t)\,.
\end{align}
The dynamics $f_k(\vec{r},t)$ is then given by
\begin{align}
\label{Eq::SUPBoltzmannFT}
\partial_t f_k+\frac{v_0}{2}\big[\partial_x (f_{k+1}+f_{k-1})-&\imag\partial_y (f_{k+1}-f_{k-1})\big]
=-\lambda(1-\e^{-(k\sigma^2)/2}) f_k+\summe{n}{-\infty}{\infty}\mathcal{I}_{n,k}f_n f_{k-n}\,,
\end{align}
where the Fourier transform of the collision integral, $\mathcal{I}_{n,k}$, has contributions coming from polar and antipolar alignments depending on the polar bias $\polarbias$:
\begin{align}
\label{Eq::collint}
\mathcal{I}_{n,k}=&\underbrace{\int\limits_{-\frac{\pi}{2}-\polarbias}^{\frac{\pi}{2}+\polarbias}\!\frac{\mathrm{d}\Delta}{2\pi}\,\mathcal{R}(|\Delta|)\left[P_\sigma\cos(\Delta(n-\frac{k}{2})-\cos(\Delta n)\right]}_{\leadsto\text{polar alignment}}
+\underbrace{\int\limits_{\frac{\pi}{2}+\polarbias}^{2\pi-\frac{\pi}{2}-\polarbias}\! \frac{\mathrm{d}\Delta}{2\pi}\,\mathcal{R}(|\Delta|)
\left[P_\sigma\cos(\frac{k\pi}{2}) \cos(\Delta(n-\frac{k}{2})-\cos(\Delta n)\right]}_{\leadsto\text{antipolar alignment}}
\, .
\end{align}
In order to find approximate stationary, spatially uniform solutions $\{f_k^{(0)}\}$ of~\eqref{Eq::SUPBoltzmannFT} we followed Ref.~\cite{Denk2016} and first calculated the uniform steady state solutions for $\{f_k^{(0)}\}$ with $k\leq k_c$ of~\eqref{Eq::SUPBoltzmannFT} setting all angular Fourier modes with $k>k_h$ as well as all spatial and temporal derivatives to zero. Depending on the global density $\bar{\rho}$ and the polar bias $\polarbias$, we find disordered solutions ($f_k^{(0)}=0,\,k=1,...k_c$), solutions with purely nematic order ($f_2k^{(0)}>0,\,f_2k-1^{(0)}=0\,k=1,...k_c$), and  solutions with polar order ($f_2k^{(0)}>0,\,f_2k-1^{(0)}>0\,k=1,...k_c$).

Next, we studied the stability of the these solutions by substituting $\rho=\bar{\rho}+\delta\rho$ and $f_k=f_k^{(0)}+\delta f_k$ with wave-like perturbations of the form
\begin{align}\label{Ansatzfouriermode}
\delta \rho(\vec{r},t) 
\sim \delta \rho_\vec{q}\,\e^{\imag \vec{q}\cdot\vec{r}}\, ,
\qquad \text{and} \qquad
\delta f_k(\vec{r},t) 
\sim \delta f_{k,\vec{q}}\,\e^{\imag \vec{q}\cdot\vec{r}}\, ,
\end{align}
where $\delta\rho_\vec{q}$ and $\delta f_{k,\vec{q}}$ are in general complex amplitudes that are assumed to be small. Periodic boundary conditions in our numeric solution impose $|\vec{q}|=n \frac{2\pi}{L},\,n\epsilon\mathbb{Z}$, where $L=\sqrt{A}$ and $A$ is the area of the (quadratic) system.
The linearized set of equations of motion for the perturbations $\delta\rho_\vec{q}(t)$, $\delta f_{k,\vec{q}}(t)$ and $\delta f_{k,\vec{q}}^*(t)$ then read 
\begin{eqnarray}\label{linstabnum}
\partial_t\delta f_k=-\frac{v_0}{2}(\nabla\delta f_{k-1}+\nabla^*\delta f_{k+1})+\lambda(\Pftnull-1)\delta f_k+\summe{n}{-\infty}{\infty}(\mathcal{I}_{n,k}+\mathcal{I}_{k-n,k})|f_{k-n}|\delta f_n\,.
\end{eqnarray}
Substituting~\eqref{Ansatzfouriermode} into Eq.~\eqref{linstabnum}, we solved the resulting linear set of equations for the maximal eigenvalue as a function of the wave vector $\vec{q}$. The real part of this eigenvalue sets the linear growth rate of the respective perturbation. A solution $\{f_k^{(0)}\}$ is linearly unstable when there is a perturbation with positive growth rate and stable otherwise. Specifically, we probed the stability of spatially uniform solutions against spatial perturbations parallel and perpendicular their order.

As summarized in Fig.~2(A), we find that for low densities, the only uniform solution is the disordered state and it is linearly stable. For larger densities we find spatially uniform solutions of purely nematic order or polar order that are stable against perturbations for arbitrary $\vec{q}$ (denoted respectively as uniform nematic and polar states in Fig.~2(A)). 
Moreover, we find regimes in which solutions with purely nematic order or polar order are stable against uniform perturbations, i.e. $|\vec{q}|=0$, but not against spatial perturbations, i.e. $|\vec{q}|>0$. For solutions with nematic order, the growth rate of perturbations in the respective regime is maximal when the perturbation is \textit{perpendicular} to the nematic order, while for solutions with polar order, the growth rate of perturbations in the respective regimes is maximal when the perturbation is \textit{parallel} to the polar order. This indicates the formation of nematic band patterns and polar wave pattern, respectively (as denoted in Fig.~2(A)). In another regime, denoted as 'bistability' in Fig.~2(A), we find a disordered solution, which is stable against arbitrary perturbations, and--at the same time--a solution with polar order, which is stable against uniform but unstable against non-uniform perturbations. This indicates the existence of both linearly stable disorder and polar wave patterns, depending on the initial conditions of the system. In Fig.~2(A),(B) we used $k_c=10$, however, choosing larger $k_c$ only leads to  negligible quantitative changes of the stability diagram Fig.~2(A) and the uniform solutions Fig.~2(B).  

\section{Numeric simulations}
\label{sec:NumSol}

In order to study the nonlinear dynamics and steady states in the kinetic Boltzmann equation in real space~\eqref{Eq::SUPBoltzmannEq} we employed the \texttt{SNAKE} algorithm as introduced in Ref.~\cite{Thuroff2013}. As tessellations we used a quadratic periodic regular lattice with periodic boundary conditions with equally sized angular slices varying from $40$ to $80$. To account for an alignment rule between the angular slices $\theta_i$ and $\theta_j$ with variable polar bias we assume polar alignment for an intermediate angle $|(\theta_i-\theta_j)|<\pi/2+\polarbias$ and antipolar alignment otherwise. For Figs.~2(B,C), we used only one lattice point in order to obtain the spatially uniform solution, while for Fig.~3 we used a $60\times60$ lattice with lattice spacing $10$. Hence, the simulated system size is $A=600\times600$. The computation time step was set to $0.3$. The system was initialized with a disordered state with small random density fluctuations around the mean density $\bar{\rho}=A^{-1}\int_{A}\rho(\vec{r},t)$. 
For densities $\bar{\rho}$ close to the onset of order we find the formation of spatial patterns: in the regime where our linear stability analysis [Fig.~2(A)] predicts polar patterns we observe traveling wave patterns as reported in Refs.~\cite{Thuroff2013} for fully polar alignment [see Fig.~\ref{fig:SNAKEpatterns}(A)]. For vanishing and small polar bias and close to $\rho_2^c$ we see the formation of nematic band patterns [see Fig.~\ref{fig:SNAKEpatterns}(B)]. The regimes of patterns and uniformly ordered solutions are in good agreement with our linear stability analysis. Similar to previous comparisons between linear stability analysis and solutions based on the \texttt{SNAKE} algorithm\cite{Denk2016}, the regimes of patterns in our numerical simulations are more restricted in parameter space than predicted by linear stability analysis [Fig.~$3$(A)], probably due to numerical diffusion in the discrete implementation of the \texttt{SNAKE} algorithm~\cite{mFahault2018outstanding}.

As detailed in the main text, for intermediate polar bias and densities, we find initially forming nematic bands which undergo an instability of polar order and eventually transform into patterns of coexisting and cycling polar and nematic symmetry (see Movie 1). We argue that this instability occurs since the density in the high density core of the nematic band exceeds the transition density $\rho_\text{nem-pol}^c$, above which polar order grows exponentially. This is shown in Fig.~\ref{fig:SNAKEtimetrace}, where we plot the local dynamics of the density, the polar and nematic order parameter within a nematic band and in the disordered area between two bands.

Depending on the system size and random seed of initial conditions, we observe that the polar instability within nematic bands can lead to different patterns with polar and nematic symmetry including replacement of nematic bands by polar waves [see Movie 2], coexistence of nematic bands and polar waves [see Movie 1] or alternating formation of nematic bands and polar waves [see Movie 3].

\section{Hydrodynamic equations}
\label{sec:HydroEqu}

In order to derive closed hydrodynamic equations from the kinetic Boltzmann equation~\eqref{Eq::SUPBoltzmannEq}, we follow~\cite{Peshkov2014} and assume that close to the onset of polar or nematic order the respective fields $f_1$ and $f_2$ as well as temporal and spatial variations are small. This assumption suggests scaling relations which allow to truncate the infinite sum in~\eqref{Eq::SUPBoltzmannFT} and get closed equations for the dominant hydrodynamic fields.   

Balancing terms in the Boltzmann equation Peshkov et al.~\cite{Peshkov2012} have proposed scaling relations for a system of polar particles with fully nematic collisions (i.e. $\polarbias=0$) according to
\begin{align}
\rho-\hat{\rho}\sim\varepsilon,\, \{f_{2k-1},f_{2k}\}_{k\geq 1}\sim\varepsilon^k,\,\partial_t\sim \varepsilon,\,\partial_{x/y}\sim\varepsilon\,.
\end{align}
With these scaling relations, one can expand the sum in~\eqref{Eq::SUPBoltzmannFT} retaining only terms of order $\varepsilon^3$ to get closed equations for the order fields  $f_{1,2,3,4}$. The equations for $f_3$ and $f_4$ yield expressions for $f_3$ and $f_4$ in terms of $f_1$ and $f_2$ and one arrives at the following hydrodynamic equations for $f_1$ and $f_2$:
\begin{subequations}
\label{Eq::SUPHydroglg}
\begin{align}
\partial_t f_1
&=
-(\alpha_0+\rho\alpha_1)f_1+\alpha_2 f_1^* f_2-\alpha_3 |f_2|^2f_1
-\frac{1}{2}(\nabla\rho+\nabla^*f_2)+\gamma_1 f_2^*\nabla f_2\label{Eq::SUPHydroglgf1}
\, , 
\\
\partial_t f_2
&= 
(-\beta_0+\rho \beta_1)f_2+\beta_2 f_1^2-\beta_3|f_2|^2f_2 -\beta'_3|f_1|^2f_2
-\frac{1}{2}\nabla f_1+\gamma_2\nabla\nabla^* f_2-\gamma_3 f_1^*\nabla f_2-\gamma_4\nabla^*(f_1 f_2)
\,,
\label{Eq::SUPHydroglgf2}
\end{align}
\end{subequations}
where $\nabla:=\partial_x+\imag\partial_y$ and the star denotes complex conjugation. The coefficients are given by 
\begin{subequations}
\label{Eq::coeff}
\begin{align}
\alpha_0&=1-P_1(\sigma)\,,\\
\alpha_1&=-\left(\mathcal{I}_{0,1}(\sigma)+\mathcal{I}_{1,1}(\sigma)\right)\,,\\
 \alpha_2&=\left(\mathcal{I}_{-1,1}(\sigma)+\mathcal{I}_{2,1}(\sigma)\right)\,, \\
\alpha_3&=-4\gamma_2\left(\mathcal{I}_{-2,1}(\sigma)+\mathcal{I}_{3,1}(\sigma)\right)\left(\mathcal{I}_{1,3}(\sigma)+\mathcal{I}_{2,3}(\sigma)\right) \\
\beta_0&=1-P_2(\sigma)\,,\\
\beta_1&=\left(\mathcal{I}_{0,2}(\sigma)+\mathcal{I}_{2,2}(\sigma)\right)\\
 \beta_2&=\mathcal{I}_{1,2}(\sigma)\,,\\
 \beta_3&=-4\gamma_2\mathcal{I}_{2,4}(\sigma)\left(\mathcal{I}_{-2,2}(\sigma)+\mathcal{I}_{4,2}(\sigma)\right)\\
 \beta'_3&=-4\gamma_2\left(\mathcal{I}_{1,3}(\sigma)+\mathcal{I}_{2,3}(\sigma)\right)\left(\mathcal{I}_{-1,2}(\sigma)+\mathcal{I}_{3,2}(\sigma)\right)\,,\\
\gamma_1&=-2\gamma_2\left(\mathcal{I}_{3,1}(\sigma)+\mathcal{I}_{-2,1}(\sigma)\right)\,,\\
\gamma_2&=1/\left(4\left(1-P_3(\sigma)-\left(\mathcal{I}_{3,3}(\sigma)+\mathcal{I}_{0,3}(\sigma)\right)\right)\right)\,,\\
\gamma_3&=2\gamma_2\left(\mathcal{I}_{-1,2}(\sigma)+\mathcal{I}_{3,2}(\sigma)\right)\,,\\
\gamma_4&=2\gamma_2\left(\mathcal{I}_{1,3}(\sigma)+\mathcal{I}_{2,3}(\sigma)\right)\,,
\end{align}
\end{subequations}
where $P_k(\sigma)=\e^{- k^2 \sigma}$ and $\mathcal{I}_{n,k}(\sigma)$ are  collision integrals defined in~\eqref{Eq::collint}. For a fully nematic collision rule, the coefficients $\alpha_0,\,\alpha_1,\,\alpha_2,\,\alpha_3,\,\beta_0,\beta_1,\,\beta_2,\,\beta'_3$ are positive. As discussed in the main text, this defines a critical density $\bar{\rho}=\beta_0/\beta_1$ above which the disordered state is unstable against growth of nematic order, whereas polar order will always decay to linear order.  

In principle, one could argue that these equations, which were derived for a fully nematic collision rule, might still be useful to study a system including a small polar bias. Indeed, the coefficient $\alpha_1$ becomes negative for larger polar bias defining a critical density at $\bar{\rho}=-\alpha_0/\alpha_1$ above which the disordered state is linearly unstable against growth of polar order. The transition densities for nematic order $\bar{\rho}=-\beta_0/\beta_1$ and polar order $\bar{\rho}=-a_0/a_1$ are in fact equivalent representations of the conditions $\mu_2(\bar{\rho},\sigma,\polarbias)=0$ and $\mu_1(\bar{\rho},\sigma,\polarbias)=0$, respectively, derived in the main text. 
We therefore included the dependence of the collision integrals on the polar bias $\polarbias$ as given by~\eqref{Eq::collint}. Even for $(\alpha_0+\alpha_1\rho)<0$ equation~\eqref{Eq::SUPHydroglgf1} might allow a polar instability when the second term, which is linear in $f_2$, dominates the first term. 

To test the validity of equations~\eqref{Eq::SUPHydroglg} in the presence of a polar bias we numerically calculated the phase diagram derived from~\eqref{Eq::SUPHydroglg} [see Fig.~\ref{fig:originalHydroPD}]. For specificity, we fixed noise value to $\sigma=0.2$. Already for zero and small polar bias we find a transition from nematic to polar order for high densities. However, this is likely an artefact from the truncation procedure which is suited for densities close to the order transition. In this high density regime, numerical solutions of equations~\eqref{Eq::SUPHydroglg} for a nematic collision rule without polar bias show unbounded growth~\cite{Peshkov2012}, indicating that higher orders neglected in the derivation of~\eqref{Eq::SUPHydroglg} become important~\cite{Peshkov2012}.
Apart from this unphysical transition to polar order for high densities at small polar bias the phase diagram of equations~\eqref{Eq::SUPHydroglg} also features a regime of polar order for larger polar bias bias. Here, similar as in Fig.~2, polar order is not restricted to densities above $\rho^c_1$ but is also present above the transition to nematic order marked by $\rho^c_2$. However, unlike the phase diagram for the kinetic Boltzmann equation [Fig.~2(A)], the phase diagram of the hydrodynamic equations~\eqref{Eq::SUPHydroglg}, Fig.~\ref{fig:originalHydroPD}, lacks a pronounced transition from a purely nematic phase to a phase of polar order. Altogether, equations~\eqref{Eq::SUPHydroglg} with the coefficients~\eqref{Eq::coeff} do not fully capture the phenomenology of a transition between nematic and polar patterns as observed in Fig.~2(A),(C).

\subsection{Simulations including higher angular Fourier modes}
\label{sec:XMDSNumSol}

The difference between the diagram in Fig.~\ref{fig:originalHydroPD} and Fig.~2(A),(C) suggest, that to capture the transformations between nematic and polar patterns as observed in Section~\ref{sec:NumSol} higher orders that have been neglected in the derivation of~\eqref{Eq::SUPHydroglg} become important. 
To test this hypothesis, we simulate the Boltzmann equation in angular Fourier space~\eqref{Eq::SUPBoltzmannFT} including the continuity equation for the density taking into account modes $f_k$ with $|k|$ up to a certain $k_h$ and setting all modes with $|k|>k_h$ and their derivatives to zero. Furthermore, we assumed periodic boundary conditions. For the numerical solution we used \texttt{XMDS2}~\cite{DENNIS2013201}, a fast Fourier transform (FFT)-based spectral solver. In the parameter regime where our simulations in real space [see Ref.~\ref{sec:NumSol}] show transformation between nematic bands and polar patterns, we find the formation of nematic bands which undergo a polar instability [see Fig.~\ref{fig:HydroDivergence}]. However, this polar instability leads to a divergence in the polar order parameter in our simulations. We obtain a similar result for numeric simulations of~\eqref{Eq::SUPBoltzmannFT} with $k_h=6$ as well as when we employ truncation assumptions other than the one detailed in the main text and Section~\ref{sec:HydroEqu} (e.g.\ when we close the truncation in~\ref{sec:HydroEqu} at $\epsilon^5$ instead of $\epsilon^3$).

\subsection{Generalized hydrodynamic equations}

For linear stability analysis and numeric simulations of the hydrodynamic equations~\eqref{Eq::SUPBoltzmannFT} together with the continuity equation [shown in Fig.4A,B] we use the parameters values 
\begin{subequations}
\label{Eq::coefffree}
\begin{align}
\alpha_0&=0.019801326693244747\,,\\
\alpha_1&= 0.4496250137624467\,, \\
\alpha_3&= 2.1897054378862726\,,\\
\beta_0&=0.07688365361336424\,,\\
\beta_1&= 0.48867097895034317\,, \\
\beta_3&= 3.007210976282877\,,\\
\beta_2&=-0.12486399707430346\,, \\
\beta'_{3}&=-0.612545745821566\,,\\
\gamma_1&=0.5258429516101398\,, \\
\gamma_2&=0.44601012183492805\,, \\
\gamma_3&=0.14709871812254954\,,\\
\gamma_4&= 1.857267053107399\,,
\end{align}
\end{subequations}
where units have been suppressed but can be easily read of from~\eqref{Eq::SUPHydroglg}. This corresponds to the values in~\eqref{Eq::coeff} for $\sigma=0.2$ and a density $\bar{\rho}=0.16$ close above the critical density $\rho_2^c$. As discussed in the main text, we now vary the nematic-polar coupling strength $\alpha_2$ and $\bar{\rho}$. Using \texttt{XMDS2}~\cite{DENNIS2013201} and a system size of $200$ on a $80\times80$ lattice, we scanned the parameter regime as indicated in Fig.~4(B) and found regimes of stable nematic band patterns, polar wave patterns, and transformations between nematic patterns due after local instabilities in the high density cores of nematic bands. Figure~\ref{fig:XMDStimetrace} shows the local density and nematic order which cross the critical nematic order above which polar order grows. In the parameter regime of transformations between nematic and polar patterns, for large system sizes we observe intriguing patterns of coexisting polar waves and nematic bands which closely interact and transform into each other in a cyclic fashion [Fig.~4(D), Movie 4], similarly as observed in Ref.~\cite{Huber2018}.

\newpage

\renewcommand{\thefigure}{S\arabic{figure}}
\setcounter{figure}{0}   

\section{Supplemental figures}
\vspace{4cm}
\begin{figure}[htp]
\centering
\includegraphics[width=0.7\textwidth]{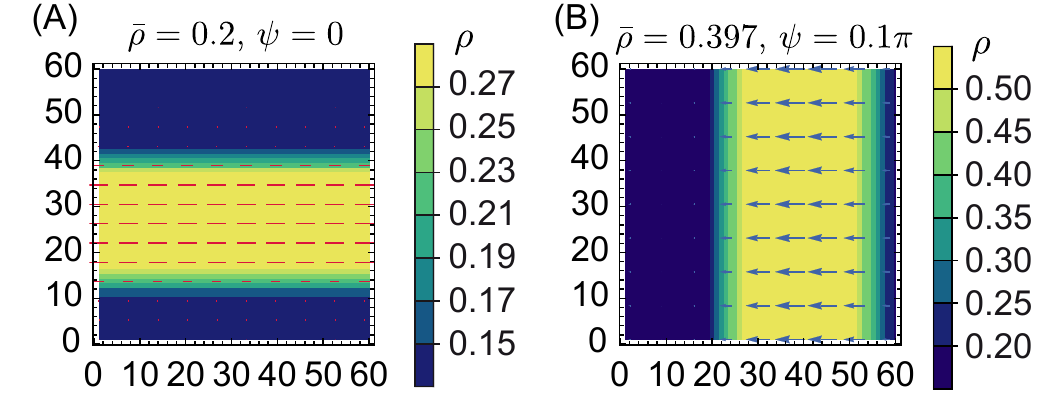}
\caption{\textbf{Spatial patterns in simulations.} For different values of $\bar{\rho}$ and polar bias $\polarbias$, we find nematic bands (A) and polar traveling waves (B), in agreement with our linear stability analysis, Fig.~2(A). The color denotes the local density, red bars and blue arrows indicated the orientation and strength of local nematic and polar order, respectively. Simulations were done on a $60\times 60$ lattice with a spacing  $10$ and $80$ angular slices. The time step was set to $0.3$.}
\label{fig:SNAKEpatterns}
\end{figure}

\newpage

\begin{figure}[htp]
\centering
\includegraphics[width=0.9\textwidth]{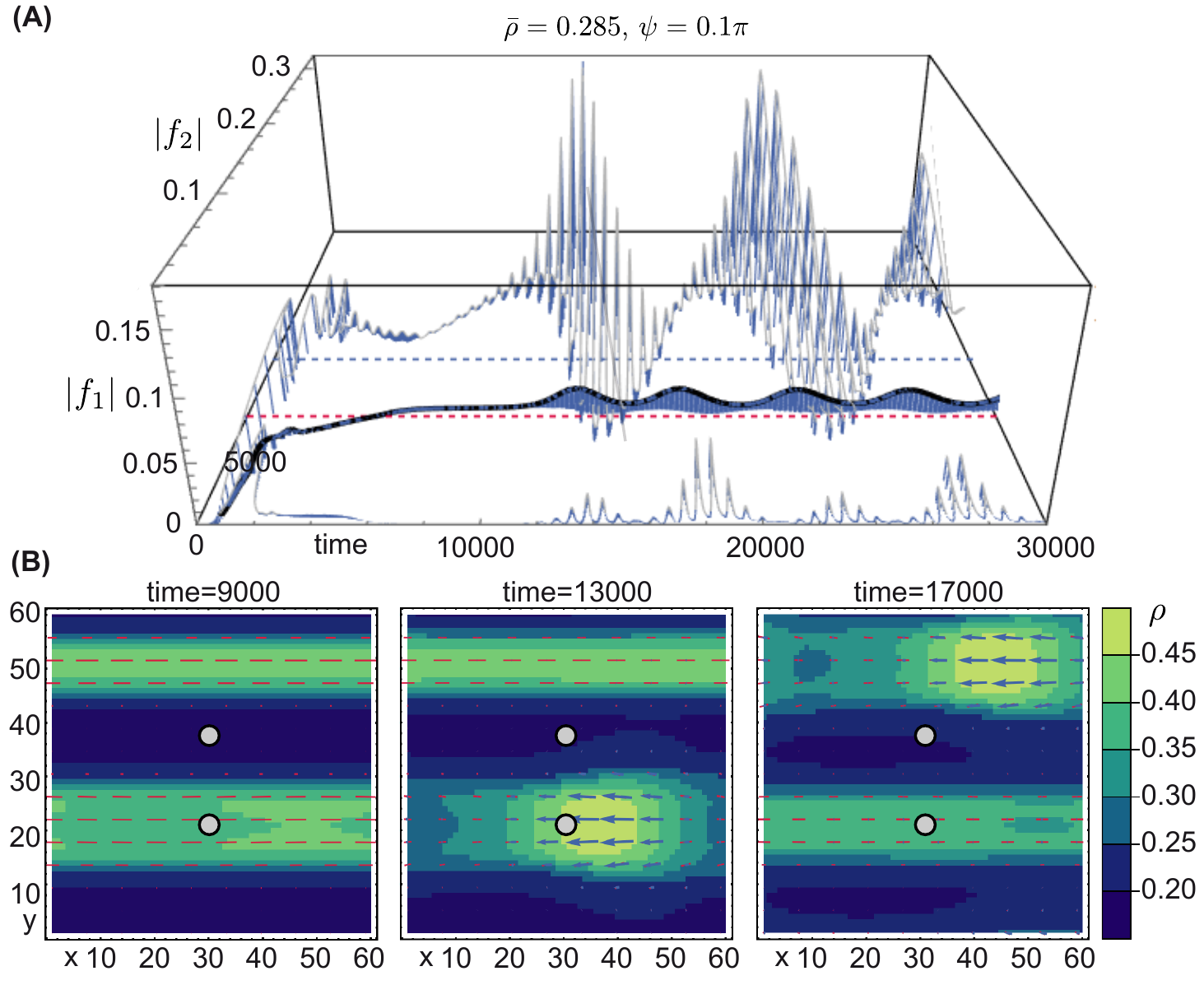}
\caption{\textbf{Polar instability in the kinetic Boltzmann approach.} \textbf{(A)} Time trace of nematic and polar order parameters (respectively $|f_2|$ and $|f_1|$) close to the core of a nematic band and in the disordered region of nematic band pattern (gray lines) as well as the average nematic and polar order (black line). Red and blue dashed lines denote the uniform steady state solution for the nematic order and the nematic order that corresponds to the critical density $\rhonempol$, respectively. After the formation of a nematic band, the nematic order within the band exceeds the value corresponding to $\rhonempol$ while in the disorder region the nematic order drops to zero. After some time polar order starts to grow within the band, which leads to the growth of global polar order. The distance between the time traces and the x-y-plane is marked by blue bars to emphasize the growth of polar order more clearly. \textbf{(B)} Snapshots at different time points before and after the polar instability shown in (A). For the respective movie, see Movie 1.}
\label{fig:SNAKEtimetrace}
\end{figure}


\begin{figure}[htp]
\centering
\includegraphics[width=0.5\textwidth]{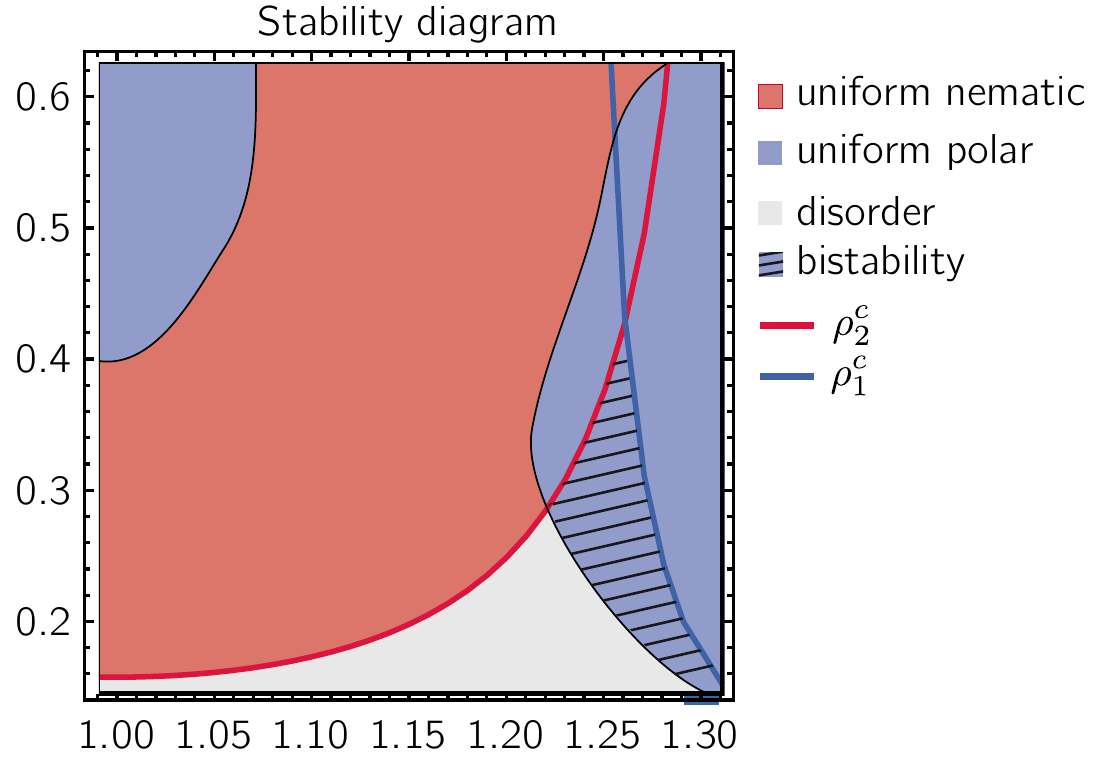}
\caption{\textbf{Stability diagram of kinetic Boltzmann-hydrodynamic equations with fully nematic truncation scheme.} The color shows different regimes in which solutions of disorder, nematic order and polar order are stable against uniform solution. Unlike the stability diagram Fig.~2(A), the phase diagram lacks a pronounced transition from nematic to polar steady states for moderate polar bias. In addition, there is an unphysical regime of polar order already for zero and small polar bias at large densities.\vspace{5cm}}
\label{fig:originalHydroPD}
\end{figure}

\begin{figure}[htp]
\centering
\includegraphics[width=0.9\textwidth]{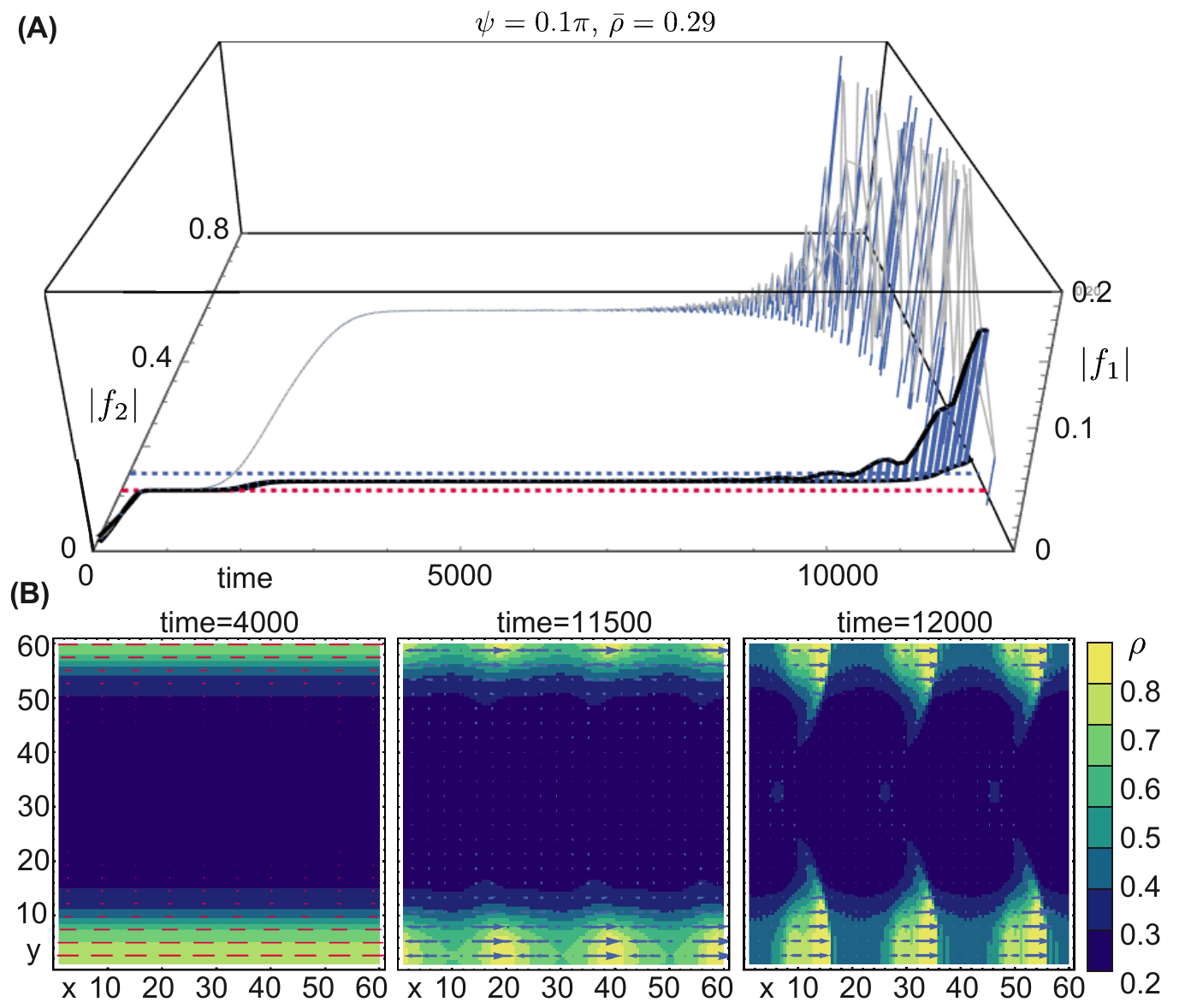}
\caption{\textbf{Polar instability in the kinetic Boltzmann equation in Fourier space.} \textbf{(A)} Evolution of the nematic and polar order of one position in the core of a nematic band (grey line) and of the spatial average (black line). The red dashed line indicates the uniform steady state solution and the blue dashed line indicates the nematic order that corresponds to the critical density $\rhonempolhydro$. The nematic order in the nematic band exceeds the value corresponding to $\rhonempolhydro$ before after some time polar order builds up. The average global polar order grows accordingly. The distance to the x-y-plane is highlighted by blue bars. Soon after the polar order grows the numeric solution diverges. \textbf{(B)} Snapshots of the patterns at different time points. The color denotes the local density. Red bars and blue arrows indicate the orientations and strengths of the local nematic and polar order, respectively. The solutions were obtained using \texttt{XMDS} software~\cite{DENNIS2013201} and using a cutoff $k_h=12$ as detailed in Section~\ref{sec:XMDSNumSol}.}
\label{fig:HydroDivergence}
\end{figure}

\begin{figure}[htp]
\centering
\includegraphics[width=0.9\textwidth]{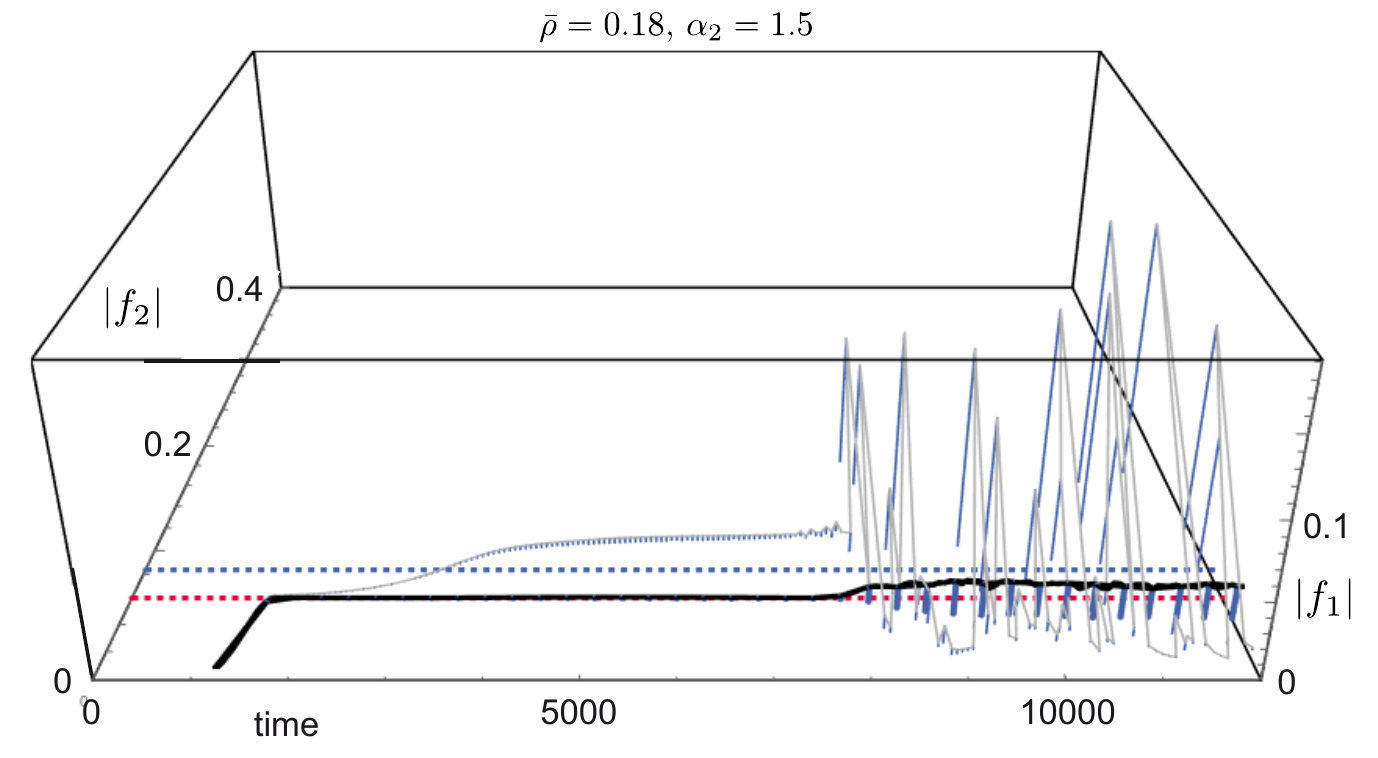}
\caption{\textbf{Polar instability in generalized hydrodynamic equations.} Evolution of the nematic and polar order of one position in the core of a nematic band (grey line) and of the spatial average (black line). The red dashed line indicates the uniform steady state solution and the blue dashed line indicates the nematic order that corresponds to the critical density $\rhonempolhydro$. The nematic order in the nematic band exceeds the value corresponding to $\rhonempolhydro$ before after some time polar order builds up. The average global polar order grows accordingly. The distance to the x-y-plane is highlighted by blue bars.\vspace{5cm}}
\label{fig:XMDStimetrace}
\end{figure}

\newpage

\section{Supplemental movie captions}
For supplemental movies, please contact the authors.

\textbf{Movie 1: Transformations between nematic bands and polar waves.} Parameters are $\bar{\rho}=0.292$ and $\polarbias=0.1\pi$. Results were obtained based on a modified version of the \texttt{SNAKE} algorithm on a $100\times 100$ lattice with spacing $8$, time step $0.3$ and $40$ angular slices. The system started from isotropic disorder with small random fluctuations with a seed different to the seed in Movies $2$ and $3$. The color shows the local density according to the color bar and the axis denote the extensions in units of the lattice spacing.

\textbf{Movie 2: Nematic bands transform into polar waves.} Parameters are $\bar{\rho}=0.285$ and $\polarbias=0.1\pi$. Results were obtained based on a modified version of the \texttt{SNAKE} algorithm on a $100\times100$ lattice with spacing $8$, time step $0.3$ and $40$ angular slices. Initial conditions were chosen uniform disordered with small random fluctuations with a seed different to the seed in Movies $1$ and $3$. The color shows the local density according to the color bar and the axis denote the extensions in units of the lattice spacing.

\textbf{Movie 3: Nematic bands transform into polar waves and back.} Parameters are $\bar{\rho}=0.285$ and $\polarbias=0.1\pi$. Results were obtained based on a modified version of the \texttt{SNAKE} algorithm on a $60\times60$ lattice with spacing $8$, time step $0.3$ and $40$ angular slices. Initial conditions were chosen uniform disordered with small random fluctuations with a seed different to the seed in Movies $1$ and $2$. The color shows the local density according to the color bar and the axis denote the extensions in units of the lattice spacing.

\textbf{Movie 4: Coexistence and transformations of nematic bands and polar waves in generalized hydrodynamic equations.} Parameters are $\bar{\rho}=0.17,\,\alpha_2=1.5$. The color denotes the local density as given in the colour bar. Simulation results were obtained with \text{XMDS} software\cite{DENNIS2013201} as detailed in~\ref{sec:XMDSNumSol} for a system of size $800\times800$ with $250$ lattice points per dimension. The simulation was initiated from a uniform disordered state with small random perturbations and the movie starts after the initial formation of nematic bands. The color shows the local density according to the color bar and the axis denote the spatial extensions.

\end{document}